\documentclass[11pt,nofootinbib,nobibnotes]{revtex4}
\usepackage{graphicx}  
\usepackage{dcolumn}   
\usepackage{bm}        
\usepackage{amssymb}   
\usepackage{latexsym}
\usepackage{dcolumn}
\usepackage{amsfonts,amssymb}
\usepackage{graphicx}
\usepackage{epsfig}
\usepackage{psfrag}
\usepackage{amsmath}
\usepackage{mathtools}

\begin{document}

\title{Curvaton: Perturbations and Reheating}

\author{Mohit Kumar Sharma}
\email{mr.mohit254@gmail.com}
\affiliation{Department of Physics \& Astrophysics, University of Delhi, New Delhi 110 007, India}

\author{Kairat Myrzakulov}
\email{kmyrzakulov@gmail.com}
\affiliation{Center for Theoretical Physics, Eurasian National University, Astana-010 008, Kazakhstan}

\author{Mudhahir Al Ajmi}
\email{mudhahir@gmail.com}
\affiliation{Department of Physics, College of Science, Sultan Qaboos University, P.O. Box 36, Al-Khodh 123, Muscat, Sultanate of Oman}

\date{\today}







\begin{abstract}
We study chaotic and runaway potential inflationary models in the curvaton scenario. In particular, we address the issue of large tensor-to-scalar ratio and red-tilted spectrum in chaotic models and reheating in runaway model in the light of latest Planck results. We show that curvaton can easily circumvent these problems and is well applicable to both type of models. For chaotic models, the observable non-Gaussianity put strong constraints on the decay epoch of curvaton as well as on its field value around the horizon exit. Besides, it can also explain the observed red-tilt in the spectrum as a consequence of its negative mass-squared value. As for the runaway inflationary models, curvaton by sudden decay into the background radiation provides an efficient reheating mechanism, whereas the inflaton rolls down from its potential and enters into the kinetic regime. To this effect, we consider the generalized exponential potential and obtain the allowed parametric space for model parameters. From the estimates on inflaton parameters, we constrained the curvaton mass and then the reheating temperature. We constrained the latter for both dominating and sub-dominating case and show that it agrees with the nucleosynthesis constraint.
\end{abstract}

\maketitle

\section{Introduction}{\label{1a}}

The latest Cosmic Microwave Background (CMB) observations \cite{ade2016planck,pl2015in,aghanim2018planck,hinshaw2013nine} has outstandingly constrained non-Gaussianity, spectral index as well as tensor-to-scalar ratio of primordial perturbations. It is generally assumed that these primordial perturbations are an artifact of quantum fluctuations of all fields present during the so called `Inflationary era'. The viable scenario of inflation can be obtained either by a single scalar-field (inflaton) \cite{linde1983chaotic,shinji-rev,trodden} or by multiple scalar-fields \cite{pertgenlang,mftye}, sources both the inflation as well as primordial perturbations. One such observationally consistent simplest multi field inflationary model is the inflaton-curvaton model in which inflaton being the dominant field drives the inflation whereas, curvaton, a sub-dominant field, is responsible for the primordial perturbations \cite{lyth2009primordial,wands2002observational,lyth-wands-2002,simplest,lyth2003primordial,bartolo-04,dimopoulos}. 

During inflation, the curvaton remains silent throughout the regime but becomes useful when it ends. In chaotic models, when inflation ends, the inflaton quickly decays into relativistic degrees of freedom, whereas the curvaton being a light scalar field initially tends to become massive. This happens due to the fact that the curvaton after inflation starts to oscillate about its value and behaves as a dust-like fluid, and as a consequence its energy density starts to increase as compared to that of the background radiation. The oscillating curvaton then also decays into radiation. If curvaton decays before dominating the total energy density of the Universe then it will lead to a significant non-Gaussianity in the power spectrum which is disfavoured by the CMB observations. Therefore, one expects curvaton to decay only after it becomes dominant \cite{langlois2003isocurvature,fonseca2012primordial,sasaki2006non}. One of the main essence of taking into account curvaton as an additional field is to reduce the tensor-to-scalar ratio $r_T$ which usually is large for single-field models and is inconsistent with observations. Although, the curvaton does not affect the tensor perturbations, it can increase the scalar perturbations and can save single field models from being ruled out \cite{fujita2014curvaton,enqvist2013mixed}. Moreover, the presence of curvaton can also explain the slight red-tilt in the scalar power spectrum as a consequence of its negative mass-squared value \cite{kobayashi2013spectator}.

On the other hand, in a runaway model, the inflaton field does not decay to background radiation but instead keeps rolling along its runaway type steep potential \cite{liddle2003curvaton,baryogenesis,cqgeng}. As a result, after a short while, its kinetic energy dominates the potential energy and the Universe enters into the so called `kinetic regime'. Since, the inflaton field potential does not have minimum, the standard reheating mechanism can not be applied here. In the literature, there exists several alternative reheating mechanisms i.e. perturbative decay \cite{reheating-wilczek}, preheating \cite{kofman1997towards}, preheating based on instant particle creation \cite{felder1999inflation} and gravitational particle production \cite{haro2020} (also see references therein). Apart from this, the reheating of the Universe can also be accomplished by the decay of the curvaton field in post-inflationary era, known as curvaton mechanism \cite{feng2003curvaton,haro2019different}. This mechanism has several advantages over others as it provides sufficiently high reheating temperature for the standard nucleosynthesis process to occur \cite{agarwal2018quintessential,hossain2015unification}. Therefore, due to its versatile behavior, curvaton is well suited to both chaotic as well as runaway models.

In this paper, we have studied both chaotic as well as runaway type models in the presence of curvaton by considering its simple quadratic potential. For both the cases, we assumed that the inflaton and curvaton are minimally coupled to each other. For chaotic inflaton potential, we have re-examined the consistency of model with latest Planck 2018 observations for quadratic as well as quartic inflaton potential. We look to determine the constraints on the mass-squared value of curvaton from the constraints on the spectral index, while assuming $r_T$ to be well within the upper-bound imposed by the observations. Also, using the non-Gaussianity observations, we constraint the decay epoch of curvaton and its field value. For runaway model, we consider the generalized exponential potential for inflaton and obtain the parametric space for by imposing the upper-bound on $r_T$. From the dominance of inflaton over curvaton during inflation, we show that the upper-bound on the curvaton mass can be expressed in terms of inflaton parameters which can then be used to estimate the reheating temperature for dominant as well as sub-dominant case. 

The layout of the paper is as follows: In section (\ref{1}), we study mixed inflaton-curvaton perturbations. For estimations, we particularly consider quadratic and quartic potentials. In section (\ref{2}), we analyze the reheating mechanism by curvaton for NO model. In both cases, we consider quadratic potential for curvaton. For numerical estimations, we take $N=55$ whenever required. 

\section{Chaotic models: Mixed perturbations} \label{1}
Let us begin with the standard chaotic potential for the slow-roll inflaton field ($\phi$) as
\begin{equation}{\label{v_phi}}
V(\phi) = \frac{k\phi^ \alpha}{\alpha M_{pl}^{\alpha-4}} \  ,
\end{equation} 
where $\alpha,k$ are dimensionless constants and $M_{pl}=2.44 \times 10^{18}$GeV is the reduced Planck mass. The parameters that describes the extent of slow-roll of the inflaton field during inflation are defined as follows:
\begin{equation}{\label{slow-roll}}
\epsilon \equiv \frac{M_{pl}^2}{2} \left(\frac{1}{V}\frac{dV}{d\phi}\right)^2, \qquad \eta_\phi \equiv M_{pl}^2\left(\frac{1}{V}\frac{d^2V}{d\phi ^2}\right) \ .
\end{equation}
In order for the inflation to take place, the condition $\epsilon, |\eta_\phi| \ll 1$ has to be satisfied. Therefore, when inflation ends, we have  
\begin{eqnarray}{\label{epsilon_end}}
\epsilon_{end} &=& \frac{M_{pl}^2}{2}\left(\frac{\alpha}{\phi_{end}}\right)^2 =1 \ , \nonumber \\ &\Rightarrow&  \phi_{end}=\frac{\alpha M_{pl}}{\sqrt{2}} \ .
\end{eqnarray}
By knowing the field value at the end of inflation i.e. $\phi_{end}$, one can find out the duration of inflationary regime or the number of e-folds $N$ as follows:
\begin{equation}{\label{N}}
N \simeq \frac{1} {M_{pl}^2} \int ^{\phi} _{\phi_{end}} \frac{V(\phi)}{V^\prime(\phi)}d\phi \simeq \frac{\phi^2}{2\alpha M_{pl}^2}-\frac{\alpha}{4} \ ,
\end{equation}
such that the slow-roll parameters $\epsilon$ and $\eta_\phi$ can be expressed in terms of $N$ as 
\begin{equation}\label{eps-eta}
\epsilon = \frac{\alpha}{4N+\alpha}, \qquad \eta_\phi = \frac{2(\alpha -1)}{4N+\alpha} \ .
\end{equation}
So far, we have considered that the inflation is being solely driven by the field $\phi$, but in general, there can be more than one field present which may not be that much important during inflation but can be in post-inflationary regime. Therefore, we assume one such field (known as curvaton ($\sigma$)) which remains sub-dominating during inflation and has a simple quadratic potential 
\begin{equation}
V(\sigma) = \frac{1}{2} m^2_\sigma \sigma ^2 \ ,
\end{equation} 
where $\sigma$ and $m_\sigma$ represents the curvaton field and its mass, respectively. During inflation, curvaton ceases to be massless $m_\sigma \ll H$ ($H$ := Hubble parameter) and behaves as a light scalar field. But after inflation ends, $H$ starts to decrease and then there comes a moment when $m_\sigma \simeq H$ i.e. curvaton becomes massive. Curvaton then starts to oscillate about its mean field value in the early stages of radiation era and as a result it behaves as a pressure-less matter component. While oscillating, its energy density $\rho_\sigma$ varies as $a^{-3}$ but that of the background radiation $\rho_\gamma$ varies as $a^{-4}$, where $a$ is the scale factor. Therefore, after some time-intervals, curvaton starts to dominate the total energy density of the Universe.

The quantum fluctuations of curvaton field during inflation converts into primordial perturbations $\xi$ (here $\xi$ is defined on the spatial slice of constant energy density) after the horizon exit ($k=aH$). This happens due to the fact that the amplitude of these quantum fluctuations gets enhanced when entered into the super-horizon regime and hence they can be treated as classical perturbations. These perturbations until they again enter inside the horizon, remains frozen. From the CMB observational constraints we know that the size of those primordial perturbations should be of the order of $10^{-5}$.
 
Apart from the adiabatic perturbations (sourced by the inflaton field), there might be some isocurvature perturbations present due to the difference between relative number densities of different components \cite{lyth2003primordial}. However, these isocurvature modes after the horizon entry convertes to adiabatic modes and disappears \cite{langlois2003isocurvature}. Therefore, only the adiabatic perturbations are relevant for the structure formation. 
For weakly interacting fields, one can write the resulting power spectrum $\mathcal{P}_\xi$ as \cite{byrnes2014comprehensive,ichikawa2008non}
\begin{equation} \label{P}
\mathcal{P}_\xi = \mathcal{P}^\phi_\xi + \mathcal{P}^\sigma_\xi = (1+\lambda)\mathcal{P}^\phi_\xi  \ ,
\end{equation}
where $\lambda \equiv \mathcal{P}^\sigma_\xi/\mathcal{P}^\phi_\xi$ is the ratio between curvaton and inflaton power spectrum. By definition, the individual power spectrum of both fields are given as
\begin{equation} \label{P_xi+P_phi}
\mathcal{P}^\phi_\xi = \frac{H_\ast ^2}{8 \pi ^2 \epsilon M_{pl}^2} \ , \qquad \mathcal{P}^\sigma_\xi  = \frac{r_d ^2H_\ast ^2}{9\pi ^2 \sigma_\ast ^2}
\end{equation}
where $H_\ast$ is the Hubble parameter, $\sigma_\ast$ is the curvaton field evaluated just before the horizon exit and $r_d \equiv \rho_\sigma/ \rho \in [0,1]$ is the ratio of curvaton energy density $\rho_\sigma$ to the total energy density $\rho$ of the Universe at the time of curvaton decay.

Now from Eqs. (\ref{P}) and (\ref{P_xi+P_phi}), we can express $\lambda$ as
\begin{equation} {\label{lambda}}
 \lambda =\frac{8}{9}\epsilon\left(\frac{M_{pl}}{\sigma_\ast}\right)^2r^2_d \ .
\end{equation}
It is evident from the above expression that for given initial conditions $\epsilon$ and $\sigma_\ast$ set by inflation, the amount of perturbations generated by curvaton is determined by when it decays i.e. on $r_d$. If it decays early (late), its contribution in density perturbations will be smaller (larger).

In the presence of curvaton, the spectral index $n_s$ and tensor-to-scalar ratio $r_T$ can be expressed as \cite{enqvist2013mixed,torrado2018measuring}
\begin{eqnarray}
n_s - 1 &\equiv &  \frac{d \ln \mathcal{P}_\xi}{d \ln k} \,= -2\epsilon + 2\eta _\sigma  - \frac{4\epsilon - 2\eta_\phi}{1 + \lambda}, \label{ns} \\
r_T &\equiv & \frac{\mathcal{P}_T}{\mathcal{P}^\sigma _\xi +\mathcal{P}^\phi _\xi} \,=\, \frac{16\epsilon}{1+\lambda}, \label{r_T}
\end{eqnarray}
where $\mathcal{P}_T$ is the tensor power spectrum and $\eta_\sigma \equiv m^2_\sigma / 3H_\ast ^2$.

Apart from the first-order field perturbations which gives rise to the power spectrum, CMB observations also encounters small but non-vanishing non-Gaussianity which dominantly comes from the quadratic term of the field perturbations and gives rise to the bispectrum. Since, the curvaton model can also gives rise to large non-Gaussianity, the model by satisfying those constraints can gives us the information about the extent of curvaton dominance required before it decays. The non-Gaussianity parameter $f_{NL}$ in terms of model parameters is written as \cite{enqvist2013mixed}
\begin{equation} \label{fnl}
\frac{6}{5}f_{NL}\! = \!\left(\frac{1}{1\!+\!\lambda}\right)^2\!\left[\frac{1}{2\epsilon}\left(1-\frac{\eta_\phi}{2\epsilon}\right)\!+ \!\lambda ^2 \left(\frac{3\!-\!4r_d\!-\!2r_d ^2}{2r_d}\right)\right] \ ,
\end{equation}
which, for lets say, $\sigma_\ast = 10^{-3}M_{pl}$ and in the limit of $r_d \to 1$ i.e. when curvaton becomes dominant, reduces to its threshold value $-5/4$. On the other hand, if curvaton decays while being subdominant, for example, for $r_d \simeq \mathcal{O}(10^{-2})$, $f_{NL}$ becomes $\mathcal{O}(10)$. Now, by making the use of stringent constraints from Planck observations, we will constraint the curvaton field parameters for quadratic ($\alpha=2$) and quartic ($\alpha=4$) form of potential.

\subsection{Planck Observational Constraints }

As we have already stated in section (\ref{1a}) that a significant contribution of curvaton in overall density perturbations can alleviate the problem of having large tensor-to-scalar ratio in the single field inflationary models with power-law potential, one can realize this from Eq. (\ref{r_T}), in which if we impose the limit $\lambda \rightarrow 0$, then it gives $r_T=0.14(0.28)$ for $\alpha = 2(4)$. Hence, from the theoretical estimates of $r_T$ it can be easily seen that the upper-bound imposed by the joint Planck TT,TE,EE+lowE+Lensing+BK14+BAO results i.e. $r_T<0.072$ \cite{aghanim2018planck} cannot be satisfied. Due to this, the single field chaotic inflationary models even by giving rise to observational consistent Gaussian adiabatic spectrum are ruled out. Therefore, in order to satisfy the observational constraint on $r_T$, $\lambda$ should be atleast $\mathcal{O} (1)$. However, one also requires $\lambda$ to be much larger than this order to get rid of the problem of having large non-Gaussianity in the spectrum which is possible if $r_d \gg \sigma_\ast/M_{pl}$.
\begin{figure}[htp!] {\label{fig1}}
\begin{minipage}[t]{0.48\linewidth}
    \includegraphics[width=\linewidth]{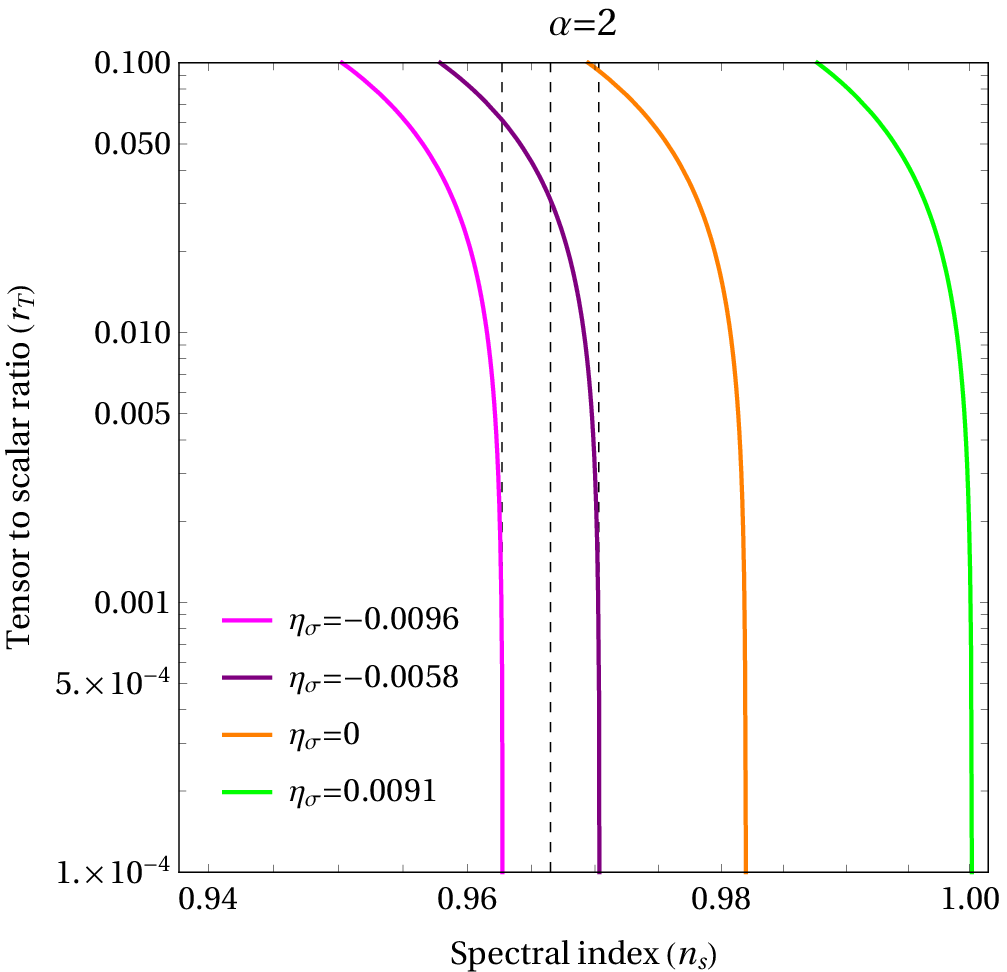}
\end{minipage}%
    \hfill%
\begin{minipage}[t]{0.48\linewidth}
    \includegraphics[width=\linewidth]{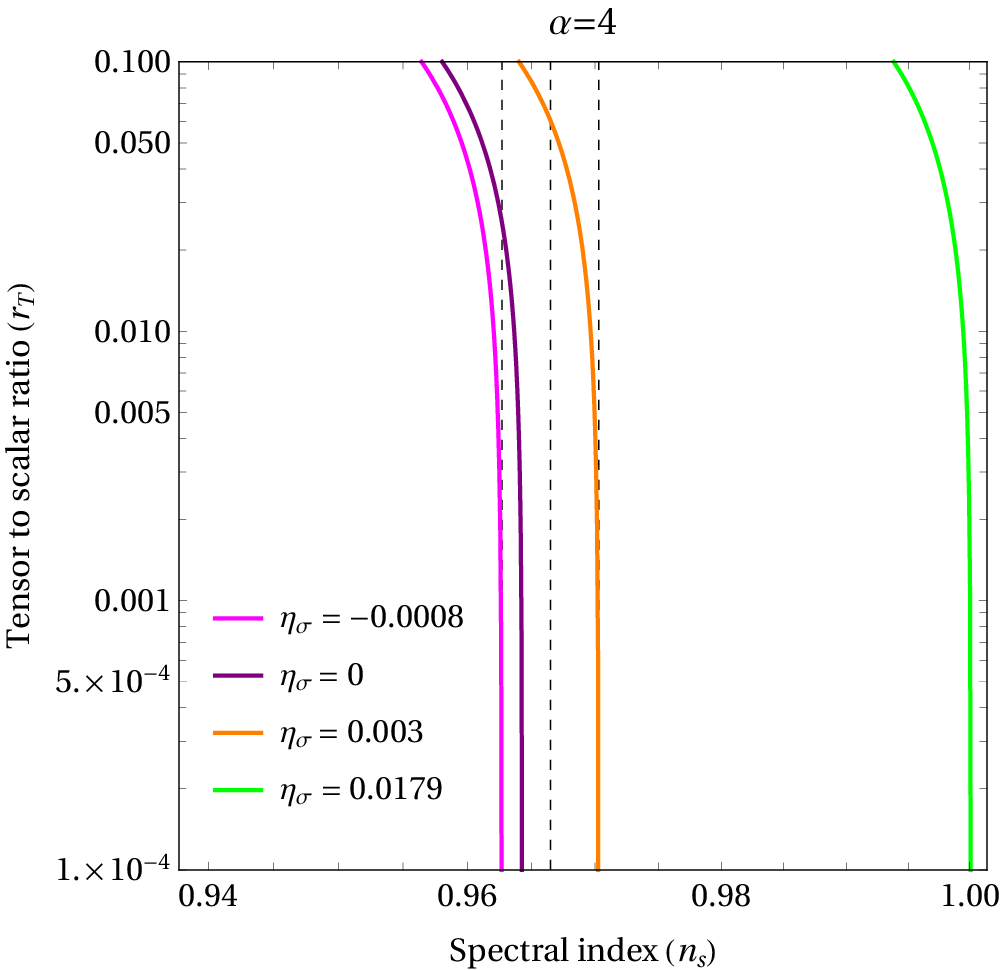}
\end{minipage} 
\caption{\footnotesize The plot between spectral index $n_s \in [0.94,1]$ and tensor to scalar ratio $r_T \in [10^{-4},0.1]$ for the quadratic [upper panel] and quartic [lower panel] inflaton potential for different values of $\eta_\sigma$. The dashed lines represents the best-fit and $1\sigma$ uncertainties in the measurements of $n_s$.  } 
\end{figure}

Now, from Eqs. (\ref{ns}) and (\ref{r_T}), by eliminating $\lambda$, we get 
\begin{equation}
r_T = 4\alpha \left( 1-n_s +2\eta_\sigma-\frac{2\alpha}{4N+\alpha}\right) \ ,
\end{equation}
where we have also used Eq. (\ref{eps-eta}). In fig. (1) we have depicted the above relation between $n_s$ and $r_T$ for quadratic and quartic potentials. In particular, we have shown the dependence of $r_T$ on $n_s$ for various values of $\eta_\sigma$, and for resasonably small value of $r_T$ one gets an estimate on $\eta_\sigma$ from the constraints on $n_s$. For example, lets say if we take $r_T =0.001$ and the $1\sigma$ confidence level of $n_s = 0.9665 \pm 0.0038$ \cite{aghanim2018planck} (shown in dashed lines), $\eta_\sigma$ for quadratic potential can only take negative values where a small positive values is still allowed for the quartic potential. We find that for $\alpha =2$, $\eta_\sigma \in [-0.0091,-0.0058]$, whereas for $\alpha=4$, $\eta_\sigma \in [-0.0008,0.003]$. It further implies that
\begin{eqnarray}
\mbox{for} \quad \alpha = 2: \quad -0.0288 \, H_\ast^2 \leq m_\sigma ^2 \leq -0.0174 \, H_\ast^2 \ ,
\end{eqnarray}
\begin{eqnarray}
\mbox{for} \quad \alpha = 4: \quad -0.0024 \,H_\ast^2 \leq m_\sigma ^2 \leq 0.009 \,  H_\ast^2 \ .
\end{eqnarray}
It suggests that to realize the red-tilt, the curvaton during inflation should inclined towards negatively curved side of the potential $V(\phi)$ (however, for quartic potential this condition can be slightly relaxed) and also its mass needs to be much less than $H_\ast$. These bounds corresponds to the fact that during inflation curvaton act as a very light scalar field and has almost negligible contribution in driving the inflation. 
Thus, apart for alleviating the problem of having large $r_T$ in single field models, the curvaton can also explain the observed red-tilted spectrum $n_s<1$ which is around $9\sigma$ level away from scale-invariance. 

\begin{figure} \label{fnlfig}
\begin{center}
\includegraphics[scale=0.45]{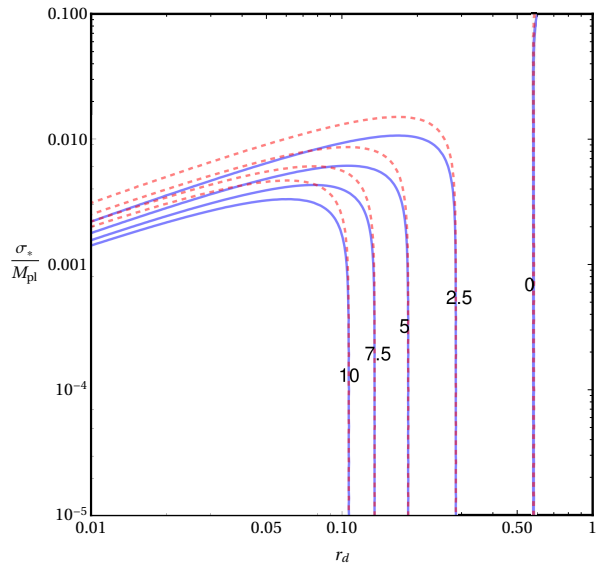} 
\caption{\footnotesize The solid and dashed contours between $r_d \in[0.01,1]$ and $\sigma_\ast/M_{pl}\in [10^{-5},0.1]$ are obtained for $\alpha=2$ and $4$, respectively. The number marked on contours represents the level of non-Gaussianity.}
\end{center}
\end{figure}

Since we are assuming that the curvaton goes through sudden decay approximation, its decay can happen either before or after dominating the energy density of the Universe. The observational quantity that can probe its decay is the non-Gaussianity parameter of the local-type $f_{NL}$ which can be expressed by using Eq. (\ref{eps-eta}) in (\ref{fnl}) as
\begin{eqnarray}{\label{f_nl}}
\frac{6}{5}f_{NL} \!= \!\left(\frac{1}{1\!+\!\lambda}\right)^2\!\left[\frac{2}{4N\!+\!\alpha}\! + \! \lambda^2\left(\frac{3\!-\!4r_d\!-\!2r_d ^2}{2r_d}\right)\right]\ .
\end{eqnarray}
In this case of mixed perturbations, if contribution of inflaton in overall perturbations is comparable to or greater than that of the curvaton, then it may lead to large non-Gaussianity in the spectrum, a signature of which is clearly absent in the Planck results. That is why in the curvaton model, curvaton has to be the dominant source for perturbations to satisfy the observational constraints, in other words, one expects curvaton to decay only after it gets dominated. Now, by replacing $\lambda$ from Eq. (\ref{lambda}) in Eq. (\ref{f_nl}), one finds that in the limit of $r_d \gg \sigma_\ast/M_{pl}$, i.e. when curvaton gets dominated, $f_{NL}$ reduces to its limiting form 
\begin{equation}
f_{NL} \simeq \frac{5}{6}\left(\frac{3-4r_d-2r_d^2}{2r_d}\right) \ .
\end{equation}
The $f_{NL}$ being a decreasing function of $r_d$ approaches to zero at $r_d=0.58$. In Fig. (2), we have depicted the corresponding dependence between $r_d$ and $\sigma_\ast/M_{pl}$ by using Eq. (\ref{f_nl}) ranges from $f_{NL} \in [0,10]$. In that figure, one can observe that if $f_{NL}$ increases, both $\sigma_\ast/M_{pl}$ and $r_d$ decreases or vice-versa. Also, for any given $f_{NL}$, $\sigma_\ast/M_{pl}$ first increases with $r_d$ and then after reaching a certain maximum limit it turns around and then decreases. Here, our goal is to find that maximum value of $\sigma_\ast$ and the corresponding range of $r_d$ by using $f_{NL}$ observations. From the Eq. (\ref{f_nl}) one can approximate $\sigma_\ast$ as
\begin{equation}
\sigma_\ast \simeq \frac{2}{9}\left[\frac{3\alpha r_d^2}{4N +\alpha }\left(\sqrt{\frac{15(3-2r_d(2+r_d))}{r_d \, f_{NL}^{local}}}-6\right)\right]^{1/2} M_{pl} \ .
\end{equation}

\begin{table}[htp!] {\label{table1}}
\centering
\renewcommand{\arraystretch}{1.4}
{
\begin{tabular}{||c||c|c||}
\hline
{\bf Observational} & \multicolumn{2}{c|}{\bf Model} \\
\cline{2-3}
{\bf Non-Gaussianity} & $\alpha=2$ & $\alpha=4$ \\
%
%
\hline\hline
1. $(f_{NL}^{local})_{_{max}}$ & \multicolumn{2}{c|}{$r_d \leq 0.179$}  \\
\cline{2-3}
(T) & $\sigma_\ast \lesssim 0.006M_{pl}$ & $\sigma_\ast \lesssim 0.008M_{pl}$ \\
\hline
2. $(f_{NL}^{local})_{_{max}}$ & \multicolumn{2}{c|}{$r_d \leq 0.206$} \\
\cline{2-3}
(T+E) &  $\sigma_\ast \lesssim 0.007 M_{pl}$ & $\sigma_\ast \lesssim 0.01 M_{pl}$ \\
\hline
\end{tabular}
}
\caption{\footnotesize Estimates of $r_d$ and $\sigma_\ast$ corresponding to the turn-around point of $\sigma_\ast$ for  $\alpha=2,4$, for the maximum value of the $1\sigma$ confidence level of $f_{NL}$ from two separate observations.}
\end{table} 

Now, from the observational constraints $f_{NL}^{local}=-0.5 \pm 5.6$ from the T-only data and $f_{NL}^{local}=-0.9 \pm 5.1$ from (T+E) data, we take the maximum $f_{NL}^{local}$ up to $1\sigma$ level $(f_{NL}^{local})_{_{max}}$ to obtain the corresponding lower limit on $\sigma_\ast$. In table (1), we have shown the range of $r_d$ and the corresponding maximum value for $\sigma_\ast$ for quadratic and quartic potentials. Since, $\sigma_\ast$ does not depend much on the choice of the inflaton field potential but for the (T+E) data we find that $\sigma_\ast$ can take values as large as $\mathcal{O}(10^{-2})M_{pl}$ for $\alpha=4$, which tends to decrease $\mathcal{P}^\sigma_\xi$. Due to this, the curvaton model seems to be more effective with quadratic inflaton potential than quartic one. One can also see that $r_d$ is independent of the choice of potential $V(\phi)$, but only depends on the non-Gaussianity parameter $f_{NL}^{local}$.

\section{Model with runaway potential: Generalized exponential potential}{\label{2}}

As we have mentioned in the Introduction, that the field with a runaway type of potential does not decay but instead keeps on rolling when inflation ends \cite{hossain2015unification,geng2017observational}. This can be realized if we consider an exponential form of potential given by
\begin{equation} \label{V}
V(\phi) = k \,   \exp\left[-\lambda \left(\frac{\phi}{M_{pl}}\right)^n\right] M_{pl}^4 \ ,
\end{equation}  
where $\lambda$ and $n$ are constants. The potential $V(\phi)$ has an interesting behaviour that during inflation it remains shallow but becomes steep in the post inflationary era \footnote{In this setup one needs to shift the field $\phi$ which is not justified in the absence of shift symmetry.}. Also at late times, it gives rise to an approximate scaling solution as $\Gamma \equiv V_{\phi\phi} V/V_\phi^2\to 1$ for large $\phi$ (for more details, see ref. \cite{geng2017observational,cqgeng}). 

The standard slow-roll parameters for this model are given as
\begin{eqnarray}
\epsilon &=& \frac{1}{2}n^2\lambda ^2\left(\frac{\phi}{M_{pl}}\right) ^{2n-2}, \label{epsilon-no} \\
\eta_\phi &=& -n\lambda\left[n-1-n\lambda\left(\frac{\phi}{M_{pl}}\right)^n\right]\left(\frac{\phi}{M_{pl}}\right)^{n-2},  \label{eta-no}
\end{eqnarray}
such that the violation of the slow-roll condition $\epsilon| _{\phi=\phi_{end}}=1$ confirms the end of the inflationary period. As a result, one can estimate field at the end of inflation i.e. $\phi_{end}$ as
\begin{equation} \label{phi-end}
\phi_{end} = \left(\frac{2}{n^2\lambda ^2}\right)^{\frac{1}{2n-2}}M_{pl} \ ,
\end{equation}
which in order to behave like a quintessence field in late-times should satisfy $\phi \gg  M_{pl}$ condition, and that can only happen if $\lambda \ll 1$.
Also, we obtain the number of e-folds $N$ as 
\begin{equation} \label{N-no}
N = \frac{1}{n\lambda(n-2)}\left[\left(\frac{\phi}{M_{pl}}\right)^{2-n} - \left(\frac{2}{n^2 \lambda ^2}\right)^{\frac{2-n}{2n-2}}\right]^{\frac{1}{2-n}} 
\end{equation}
which is valid for any $n>1$ except $n=2$, as there exist a singularity. Now, by re-expressing the above equation in terms of $\phi$ as
\begin{equation} {\label{phi-no}}
\phi = \left[n(n-2)\lambda N + \left(\frac{2}{n^2\lambda ^2}\right)^{\frac{2-n}{2n-2}}\right]^{\frac{1}{2-n}}M_{pl} \ ,
\end{equation}
we obtain a simplified expression by imposing the large-field limit for the reason mentioned earlier
\begin{equation} {\label{phi-no-lfl}}
\phi \simeq \left[n(n-2)\lambda N \right]^{\frac{1}{2-n}}M_{pl} \ .
\end{equation}

By using the constraint $r_T =16\epsilon  < 0.07$ \cite{aghanim2018planck}, we obtain the parametric space between $n$ and $\lambda$ (see Fig. (3)) in which the shaded region represents the allowed parametric space whereas the white portion is excluded. Since, we have already stated before that the large-field limit demands $\lambda \ll 1$, it is clear from Fig. (3) that to satisfy this condition $n$ must be greater than unity (except $n \neq 2$). So in order to give rise to quintessential effects at late times, exponential potential with $n>1$ is favoured over $n=1$. 

Now, to estimate $\lambda$ for each $n$, let us consider the standard expression of spectral index $n_s =1 -6\epsilon+ 2\eta_\phi$, which can be written in more explicit form by using Eq. (\ref{epsilon-no}) and (\ref{eta-no}) as
\begin{equation}
n_s-1 = -n\lambda \left[2(n-1)+n\lambda\left(\frac{\phi}{M_{pl}}\right)^n\right]\left(\frac{\phi}{M_{pl}}\right)^{n-2}  \ .
\end{equation}
By again making use of the best fit value $n_s=0.9665$ (Planck TT,TE,EE+lowE+lensing+BAO 2018), we obtain $\lambda = 5.6 \times 10^{-6}$,  $2.8\times 10^{-10}$ and $7.8\times 10^{-15}$ for $n=4,6$ and $8$, respectively. Also, by considering the COBE normalization \cite{cobe} i.e. $k^{1/4}=0.013 \, r_T^{1/4}$ together with Eqs. (\ref{epsilon-no}) and(\ref{phi-no}), we obtain
\begin{eqnarray} {\label{k-no}}
k &\simeq& 2.8 \times 10^{-8} \, \left(n^2 \lambda ^2 [n\lambda(n-2)N]^{\frac{2(n-1)}{2-n}}\right) \nonumber \\ &\simeq& 9.5 \times 10^{-10}\ ,
\end{eqnarray}
for all obtained sets of $n$ and $\lambda$. By using these estimations, we will now constrain the reheating temperature.
\begin{figure}[!htp]
\centering
\includegraphics[scale=0.52]{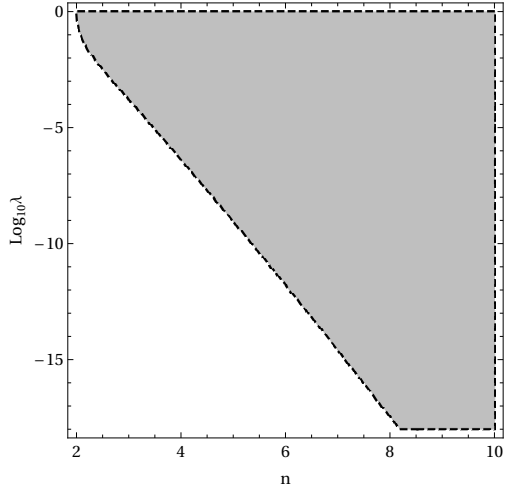}
\caption {\it The allowed parametric space between $n$ and $\log_{10}\lambda$, obtained using the constraint $r_T<0.07$. }
\end{figure}

\subsection{Constraints on Reheating Temperature }

As we know that in runaway models, the inflaton field does not decay and therefore an alternative source to execute the reheating mechanism is required. For this purpose, one requires another light scalar field like curvaton $m_\sigma \ll H_{inf}$, which can take care of reheating process. After inflation ends, the Universe enters into the kinetic regime and curvaton starts to oscillate about its mean field value and finally becomes massive $m_\sigma \simeq H_{end} $. But in order to prevent another inflationary scenario, curvaton still remains sub-dominant in the beginning of kinetic regime, by satisfying the following condition
\begin{eqnarray}  {\label{sigma<<}}
\rho_\sigma &\ll \rho_\phi = 3H^2_{end}M_{pl}^2 \quad \Rightarrow \quad
\sigma_{\ast}^2 &\ll \frac{3}{4\pi}M_{pl}^2 \ ,
\end{eqnarray}
where $\rho_\sigma \simeq m^2 \sigma_\ast^2$. Here, we have assumed that  $ \sigma_\ast \simeq \sigma_{i} $ (where $\sigma_i$ is the initial field value). Also, the sub-dominant condition for curvaton during inflation constraints the curvaton mass as
\begin{eqnarray} {\label{m<<}}
&&\frac{V(\sigma)}{V(\phi)} = \frac{1}{2}\frac{m^2_\sigma \sigma _\ast ^2}{k \,\, \exp[-\lambda (\frac{\phi}{M_pl})^n]M_{pl}^4} \ll 1 \nonumber\\
&&\Rightarrow m_\sigma \ll \left(\frac{8\pi k \,\, \exp[-\lambda (\frac{\phi}{M_pl})^n]}{3}\right)^{1/2}M_{pl},
\end{eqnarray} 
where we have used Eq. (\ref{sigma<<}). Now for the obtained values of $n$, $\lambda$ and $k$, we obtain the upper bound on $m_\sigma$ as
\begin{equation}
m_\sigma \ll 2.89 \times 10^{-5}M_{pl},
\end{equation}
which fulfills the above said requirement $m_\sigma  \ll  H_{inf}$, as $H_{inf} \leq 10^{-6}M_{pl}$.
  
As we have stated above that curvaton by sudden decay, creates all the matter present in the Universe. Therefore, by using the standard definition of decay parameter $\Gamma_\sigma$, we can constrain its decay epoch. Let us consider both the cases for curvaton decay i.e. dominating and sub dominating.

For dominating case when $\rho_\sigma > \rho_\phi$, $\Gamma_\sigma$ satisfies the following condition \cite{feng2003curvaton}
\begin{equation}{\label{gamma}}
\frac{\Gamma_\sigma}{m_\sigma} \leq \frac{\sigma_\ast^2}{3M_{pl}^2} < 1
\end{equation}
and its corresponding reheating temperature, $T_{rh} \sim \rho^{1/4}_\sigma \sim \sqrt{\sqrt{3}M_{pl}\Gamma_\sigma}$ \cite{haro2019different} which from Eq. ({\ref{gamma}}) can shown to be bounded as
\begin{equation} \label{T_rhdom}
T_{rh} \leq \sqrt{\frac{m_\sigma \sigma_\ast ^2}{\sqrt{3}M_{pl}}} \ .
\end{equation}
Also, in order to avoid the large production of gravitational waves, the signature of which is clearly absent in the CMB observations, it is necessary to consider the constraint imposed by Big Bang Nucleosynthesis (BBN) on the model parameters. In particular, the ratio between the energy density of massless particles to the background energy density in kinetic regime, also known as the heating efficiency $\Theta\equiv \frac{m_\sigma^2 \sigma_\ast^2}{3 H_{kin}^2 M_{pl}^2}$ is constrained as \cite{haro2019different}
\begin{equation}
\Theta \geq 4.64 \times 10^{-4} \left(\frac{H_{kin}}{M_{pl}}\right)^3 \ .
\end{equation}
Let us take $H_{kin} \simeq 10^{-6}M_{pl}$, which implies $m_\sigma \sigma_\ast \geq 3.734\times 10^{-17}M_{pl}^2$. Also, since $1MeV \leq T_{rh}\leq 10^{9}$GeV, we find that $m_\sigma \sigma_\ast^2 \leq 3 \times 10^{-19}M_{pl}^3$ from Eq. (\ref{T_rhdom}). In fig. (4), we plot the allowed parametric region between $m_\sigma$ and $\sigma_\ast$ by using the above constraints. Note that both constraints can be satisfied simultaneously if 
\begin{equation}
\sigma_\ast \leq 8.9\times 10^{-3}M_{pl}  \quad \mbox{and} \quad  m_\sigma \geq 3.78 \times 10^{-15}M_{pl} \ .
\end{equation}
To estimate $T_{rh}$, let us safely consider $m_\sigma =10^{-8}M_{pl}$ and $\sigma_\ast = 10^{-5}M_{pl}$ in Eq. (\ref{T_rhdom}), which gives
\begin{equation}
T_{rh} \leq 7.59 \times 10^{-10} M_{pl} \ .
\end{equation}
\begin{figure}[t!] {\label{sigma-m}}
\centering
\includegraphics[width=2.5in]{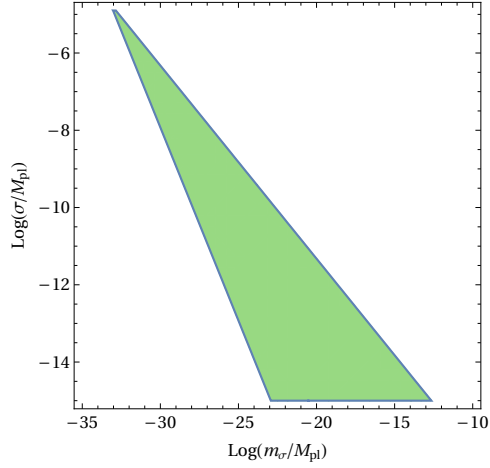} 
\caption{\footnotesize The plot between $m_\sigma/M_{pl}$ and $\sigma_\ast/M_{pl}$ obtained by imposing the constraints on $T_{rh}$ as well as heating efficiency. The allowed region is shown in color. } 
\end{figure}

Similarly, if curvaton decays while being sub-dominant i.e. $\rho_\sigma < \rho_\phi$, $\Gamma_\sigma$ satisfies
\begin{equation}{\label{sub-gamma}}
\frac{m_\sigma \sigma_\ast ^2}{3 M_{pl}^2} \leq \Gamma_\sigma \leq m_\sigma \ .
\end{equation}
Assuming that reheating happens instantaneously, the curvaton can decay when the scale factor at the moment of reheating $a_{rh}$ satisfies $a_{osc}<a_{rh}<a_{eq}$, where $a_{osc}$ and $a_{eq}$ are the scale factors when curvaton oscillates and at the equality epoch, respectively. The reheating temperature in this case is given as
\begin{equation}{\label{sub-T}}
T_{rh} \simeq \sqrt{\frac{m_\sigma^{3/2} \sigma_\ast^{3}}{3 M_{pl}^2 \Gamma_\sigma^{1/2}}} \ .
\end{equation}
Rearranging and plugging back in Eq. (\ref{sub-gamma}), we get
\begin{equation}
\sqrt{\frac{m_\sigma \sigma_\ast^3}{3 M_{pl}^2}} \leq T_{rh} \leq \sqrt{\frac{m_\sigma \sigma_\ast^2}{\sqrt{3}M_{pl}}} \ ,
\end{equation}
taking again the same values $m_\sigma=10^{-8} M_{pl}$ and $\sigma_\ast=10^{-5}M _{pl}$, we obtain
\begin{equation}
1.82 \times 10^{-12}M_{pl} \leq T_{rh} \leq 7.59 \times 10^{-10} M_{pl} \ ,
\end{equation}
which is, as expected, well satisfies the requirement for the standard BBN process to occur. Note that in this case one can satisfy the BBN constraint for a wide range of $\sigma_\ast$ and $m_\sigma$.

\section{Discussion and Conclusions}
In this paper, we have examined the viability of both chaotic as well as runaway inflationary models. For chaotic one, we have carried out analysis for two types of potential namely, quadratic and quartic. We have shown that for both forms of potentials the presence of curvaton field can indeed alleviate the problem of having large tensor-to-scalar ratio specific to the single field inflationary models. We have also constrained $\eta_\sigma$ by using the current observational constraints on $n_s$, and found that it can be always negative for quadratic potential but can also take small positive values for quartic potential. We found that the maximum mass-squared value of curvaton for the quartic potential is around one order of magnitude smaller than quadratic. Moreover, we also obtain upper-bound on $r_d$ and $\sigma_\ast$ from the maximum $1\sigma$ observational limit on the local non-Gaussianity parameter.

As for the runaway models, which are characterized by a run away type potential, the inflaton field survives to account for late time physics. We have thus considered the generalized exponential potential which can successfully account for inflation. 
After inflation, the field potential becomes steep and despite the fact it is not exponential, it might give rise to scaling behaviour in the asymptotic regime as $\Gamma \equiv V_{\phi\phi} V/V_\phi^2\to 1$ for large values of the field. 
In this case, one could use an alternative reheating mechanism based on the curvaton decay, which interacts with inflaton only gravitationally. In this paper, we have explored that curvaton reheating which seems to be an ideal in this case. 

As for the parameter estimation, we have again considered Planck 2018 results and have depicted the parametric space between $n$ and $\lambda$ and obtain their possible set of values. From the viable possible values of both $n$ and $\lambda$, we have estimated parameter $k \simeq 9.5 \times  10^{-10}$ and also obtained upper bound on $m_\sigma$ which again confirms that even in this case curvaton has to be very light scalar field. We have also obtained the allowed limits for $T_{rh}$ for dominating as well as sub-dominating case which satisfies the BBN constraints.

We have thus demonstrated that curvaton scenario is appropriate as well as advantageous to both chaotic as well as runaway type models.

\section*{Acknowledgements}
We thank M. Sami for useful discussions. MKS acknowledges the financial support by the Council of Scientific and Industrial Research (CSIR), Government of India. M. Al Ajmi  is supported by Sultan Qaboos University under the Internal Grant (IG/SCI/PHYS/19/02).

\end{document}